\documentclass[preprint,pre,aps]{revtex4-1}
\usepackage[utf8]{inputenc}
\usepackage{graphics,graphicx,amssymb,amsmath,dcolumn,bm,url,amsbsy,pgf}

\begin{document}

\title{Transport barriers in symplectic maps}

\author{R. L. Viana $^1$ \footnote{Corresponding author. e-mail: viana@fisica.ufpr.br}, I. L. Caldas $^2$, J. D. Szezech Jr. $^3$, A. M. Batista $^3$, C. V. Abud $^4$, A. B. Schelin $^5$, M. Mugnaine $^1$, M. S. Santos $^2$, B. B. Leal $^2$, B. Bartoloni $^2$,  A. C. Mathias $^1$, J. V. Gomes $^1$, and P. J. Morrison $^6$}
\affiliation{$^1$ Departamento de F\'{\i}sica, Universidade Federal do Paran\'a, Curitiba, Paran\'a, Brazil; $^2$ Departamento de F\'{\i}sica Aplicada, Instituto de F\'{\i}sica, Universidade de S\~ao Paulo, S\~ao Paulo, S\~ao Paulo, Brazil; $^3$ Departamento de Matem\'atica e Estat\'{\i}stica, Universidade Estadual de Ponta Grossa, Ponta Grossa, Paran\'a, Brazil; $^4$ Universidade Federal de Goi\'as, Goi\'as, Brazil; $^5$ Instituto de F\'{\i}sica, Universidade de Bras\'ilia, Brasília, Distrito Federal, Brazil; $^6$ Department of Physics, The University of Texas at Austin, Texas, United States}

\date{\today}

\begin{abstract}
 Chaotic transport is a subject of paramount importance in a variety of problems in plasma physics, specially those related to anomalous transport and turbulence. On the other hand, a great deal of information on chaotic transport can be obtained from simple dynamical systems like two-dimensional area-preserving (symplectic) maps, where powerful mathematical results like KAM theory are available. In this work we review recent works on transport barriers in area-preserving maps, focusing on systems which do not obey the so-called twist property. For such systems KAM theory no longer holds everywhere and novel dynamical features show up as non-resistive reconnection, shearless curves and shearless bifurcations. After presenting some general features using a standard nontwist mapping, we consider magnetic field line maps for magnetically confined plasmas in tokamaks.
\end{abstract}

\maketitle 

\section{Introduction}

The main goal of the study of transport in Hamiltonian systems is to characterize the motion of groups of trajectories from one region of phase space to another \cite{meiss05}. When dealing with non-integrable Hamiltonian systems, the study of transport is complicated by the coexistence of periodic, quasi-periodic, and chaotic orbits \cite{mei}. In particular, chaotic transport is an issue of major importance in plasma physics, since plasma turbulence is the ultimate cause of anomalous transport in magnetically confined plasmas \cite{wootton}. 

Fortunately many features of chaotic transport observed in real plasmas are also present in low-dimensional systems like area-preserving symplectic maps \cite{balescu}. If the latter satisfy the so-called twist condition, the celebrated KAM theorem warrants the existence of invariant tori with sufficiently irrational rotation numbers, provided the perturbation strength is small enough \cite{lichtenberg}. KAM tori, or invariant curves, act as dikes preventing transport in a large scale in phase space. As the perturbation strength is increased, however, these tori are progressively destroyed, leaving there cantori as their remnants \cite{mmp}. 

If, however, the twist condition does not hold everywhere in the phase space region of interest, KAM theory no longer applies everywhere in phase space. As a consequence, novel  features show up that influence transport in a dramatic way. For example, there are shearless tori for which the rotation number has a local extreme. These shearless tori, even after their breakup, if the perturbation is strong enough, decreases transport in such a way that it becomes an effective transport barrier \cite{morrison}. One of the observable consequences of these barriers is a ratchet current, when there is a symmetry breaking  \cite{novo2}. We have recently investigated the effect of a weak dissipation in nontwist systems, with the formation of shearless attractors \cite{novo1}.

Nontwist systems appear in several problems of plasma physical interest, like magnetic field line structure in Tokamaks with reversed magnetic shear \cite{del,oda,balescu2}, the ${\bf E} \times {\bf B}$ drift motion of charged particles in a magnetic field \cite{horton,del1}, transport by traveling waves in shear flows with non-monotonic velocity profiles \cite{del2}, laser-plasma coupling \cite{lp}, and magnetic field structure in double tearing modes \cite{stix}, among others.

The so-called standard nontwist map (SNTM), proposed by Morrison and Del Castillo-Negrete in 1993, is considered a paradigm symplectic map for theoretical and computational investigations of nontwist systems \cite{del2}. Drift trajectories in Tokamaks with reversed electric shear can be reduced to the SNTM \cite{horton}. The SNTM is also obtained for the magnetic field line behavior in tokamak, when the safety factor radial profile of the magnetic flux surfaces is non-monotonic, having local extrema \cite{portela}.

Effective transport barriers observed in nontwist symplectic maps can help to understand the formation of internal transport barriers in tokamak plasmas. The latter are produced by modifications of the current, safety factor, or electric field profiles by using external heating and current drive \cite{wolf} or voltage biasing \cite{toufen}. Internal transport barriers can provide high tokamak confinement at modest plasma current values \cite{challis}. 

In this work we aim to review recent theoretical and computational works aiming to understand the formation of effective transport barriers in nontwist symplectic maps, having in mind applications in plasma physics problems such as the magnetic field line structure with reversed shear. The basic dynamical mechanism underlying the formation of such barriers is the formation of dimerized magnetic island chains in both sides of the shearless curve, due to the non-monotonicity of the winding number profile. As the non-integrable perturbation is strong enough, when these islands overlap they are progressively destroyed, leaving in their places internal transport barriers that reduce (while not blocking at all) chaotic transport through them. 

The rest of the paper is organized as follows: in Section II we review some basic properties of the standard nontwist map, focusing on the existence and destruction of the shearless curve and the formation of internal transport barriers. Section III considers two magnetic field line maps in tokamaks with nontwist properties and the consequent formation of transport barriers. Section IV presents the newly discovered phenomenon of shearless bifurcation, and also the effect of reversed current. The last Section contains our Conclusions. 

\section{Standard nontwist map}

Let us consider a two-dimensional area-preserving map of the general form 
\begin{align}
 \label{gy}
 y_{n+1} & = y_n - f(x_n), \\
 \label{gx}
 x_{n+1} & = x_n + \omega(y_{n+1}), \qquad (\mbox{\rm mod } \, 1),
\end{align}
where $y_n \in \mathbb{R}$ and $x_n \in [0,1)$ are canonical variables. We require that the function $f$ be periodic with period-$1$. If the latter vanishes everywhere in the cylindrical phase space we have simply $x_{n+1} = x_n + \omega(y_n)$ and the system is integrable. Hence the orbits lie on curves $y_n = const.$, along which $\omega(y_n)$ is the so-called rotation number, defined more generally as 
\begin{equation}
 \label{winding}
 \omega = \lim_{n\rightarrow\infty} \frac{x_n - x_0}{n}.
\end{equation}

The system (\ref{gy})-(\ref{gx}) is a twist map, provided the condition
\begin{equation}
 \label{twist}
 \left\vert \frac{\partial x_{n+1}}{\partial y_n} \right\vert = \vert \omega'(y_{n+1}) \vert \ge c > 0,
\end{equation}
holds for every value of $(y_n,x_n)$ along a curve in the phase plane. A nontwist system is such that this condition is not fulfilled somewhere, for example when the derivative above (also called shear) crosses zero. 

\begin{figure}
\begin{center}
\includegraphics[scale=0.1]{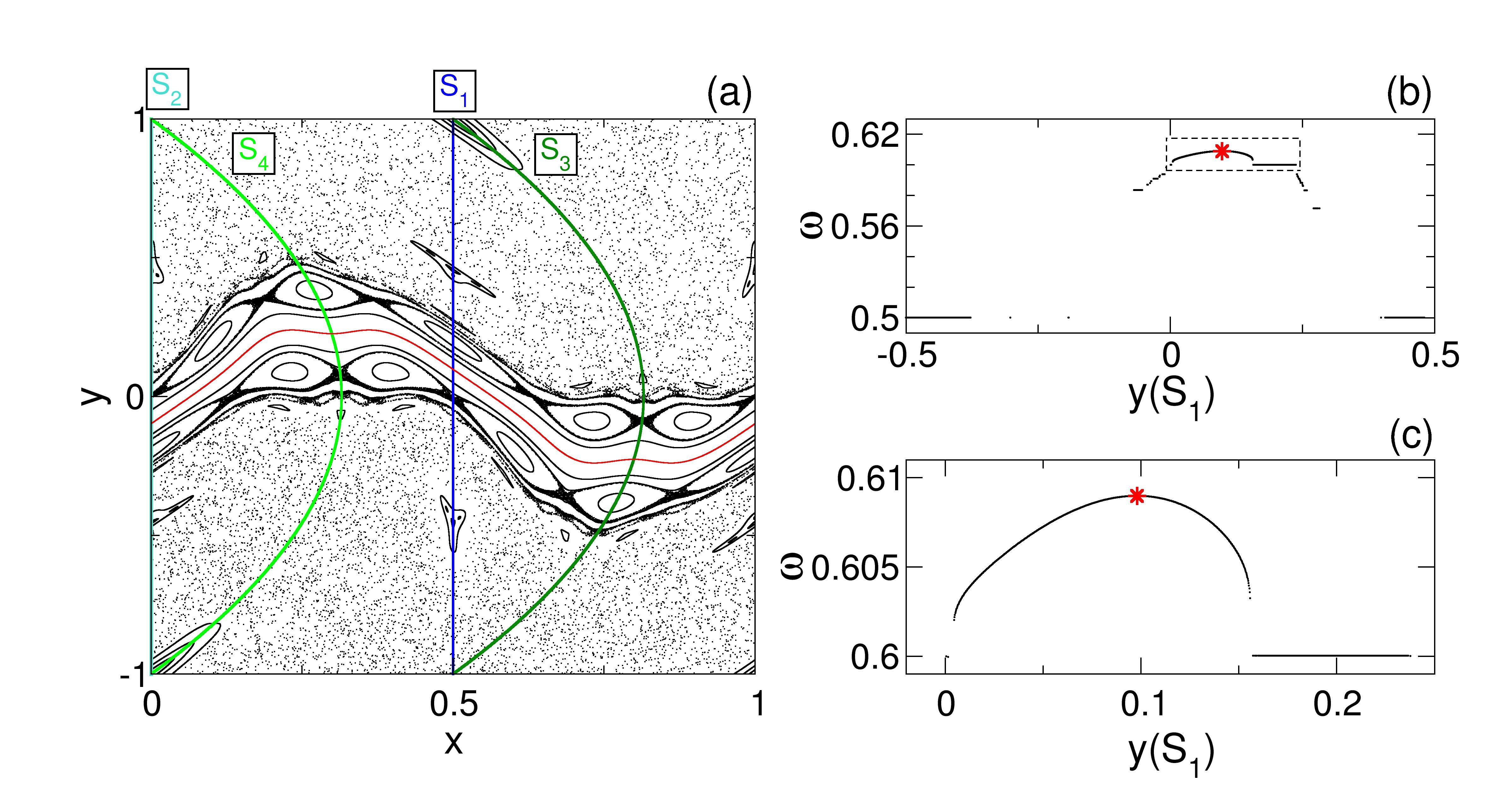}
\caption{(color online) (a) Phase space  of the SNTM for $a = 0.631$ and $b = 0.475$. The shearless curve is represented in red, and we also display the four symmetry lines. (b) Rotation number profile along the symmetry line $S_1: x = 0.5$. (c) a zoom of a region of (b). The red star indicates the $y$-position of the shearless curve along $S_1$.}
\label{fig1}
\end{center}
\end{figure}

\begin{figure}
\begin{center}
\includegraphics[scale=0.1]{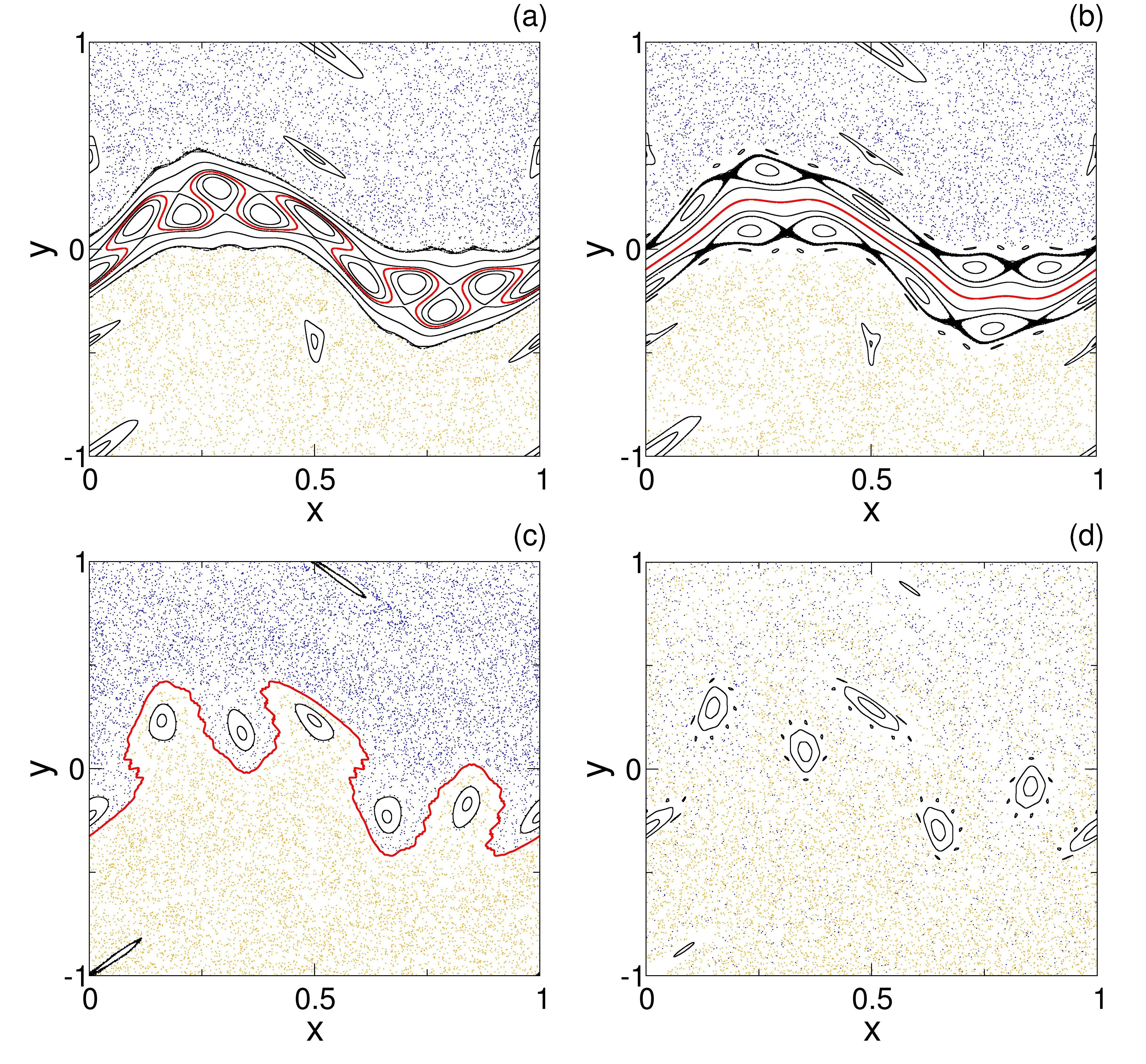}
\caption{(color online) Phase space of the SNTM for $b = 0.475$ and (a) $a = 0.623$, (b) $a = 0.631$, (c) $a = 0.6986$, (d) $a = 0.71$. The shearless curves are indicated by the red curves. }
\label{fig2}
\end{center}
\end{figure}

A two-dimensional paradigm for the study of nontwist systems is the so-called standard nontwist map (SNTM), for which $f(x) = - b\sin(2\pi x)$ and $\omega(y) = a(1-y^2)$\cite{del2}:
\begin{align}
 \label{sntmy}
 y_{n+1} & = y_n - b \sin{(2 \pi x_n)}, \\
 \label{sntmx}
 x_{n+1} & = x_n + a (1-y^2_{n+1}), \qquad (\mbox{\rm mod } \,1),
\end{align}
where $a \in [0,1)$ and $b\in \mathbb{R}$. The parameter $b$ is a measure of the non-integrability of the system, and $a$ is proportional to the shear along $(x,y)$ curves. This map has for symmetry lines, namely $S_1 = \{(x,y)|x=1/2\}$, $S_2 = \{(x,y)|x=0\}$, $S_3 = \{(x,y)|x=a(1-y^2)/2\}$, $S_4 = \{(x,y)|x=a(1-y^2)/2+1/2\}$, that are useful to find periodic orbits of any period \cite{del3}. For example, orbits with odd period $n$ on $S_2$ are obtained by searching for points $(x=0,y)$ on $S_2$ that are mapped to $S_3$ or $S_4$ after $(n+1)/2$ iterations, which reduces to a one-dimensional root finding problem \cite{del3}.

\begin{figure}
\begin{center}
\includegraphics[scale=0.1]{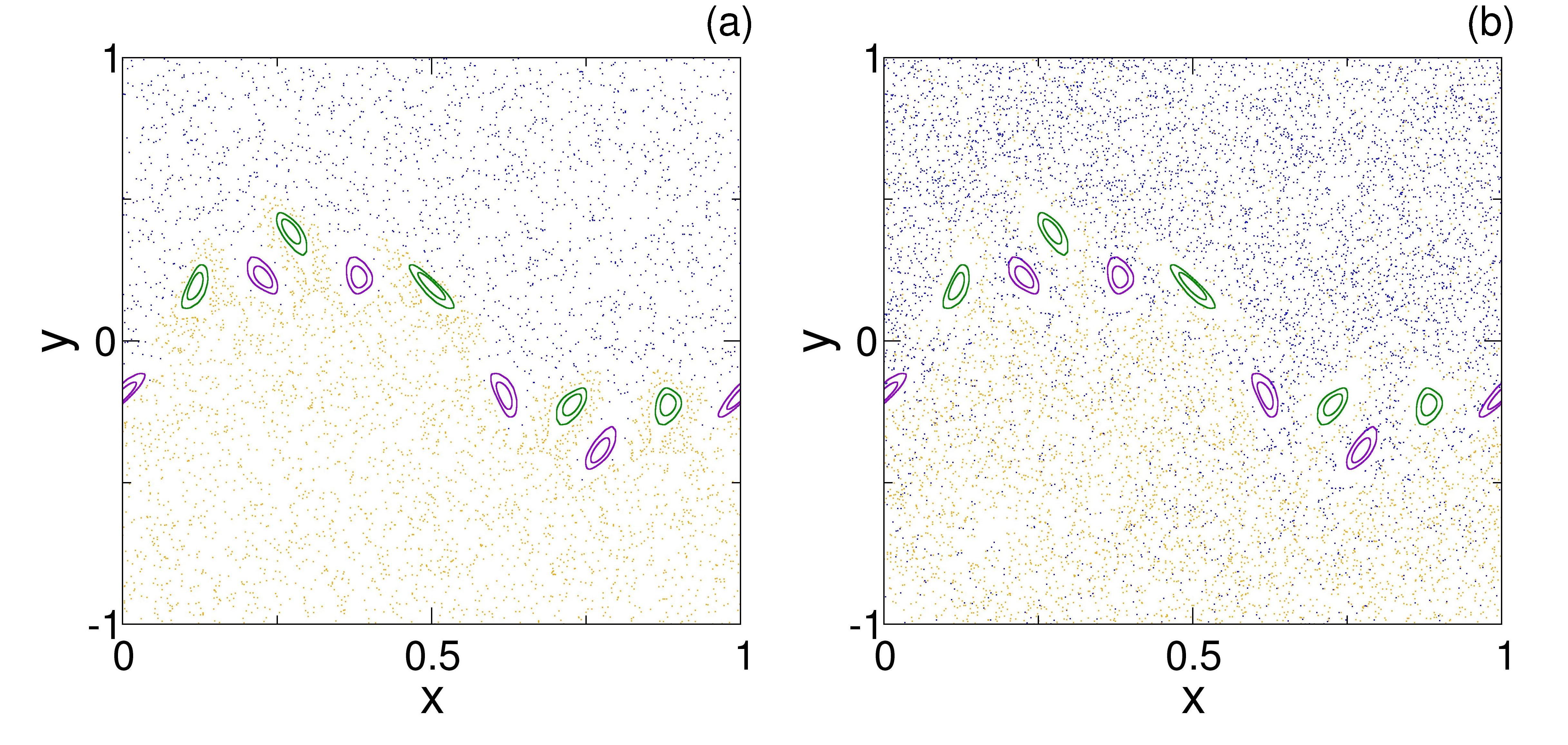}
\caption{(color online)  Phase space of the SNTM for $b = 0.619$, (a) $a = 0.642$, (b) $a = 0.645$.}
\label{fig3}
\end{center}
\end{figure}

If $b=0$ the SNTM is integrable and the shear is simply $\omega'=2ay$, changing sign at $y = 0$, for which the map is nontwist. At each side of $y = 0$ we have two invariant curves $y = y_0$ and $y = - y_0$ with the same rotation number for a given value of $y_0$. Switching on the perturbation ($b \ne 0$)  there will appear twin chains of periodic islands with the same rotation number, as illustrated by Fig. \ref{fig1}(a), where a phase portrait of the SNTM is shown for $a = 0.631$ and $b = 0.475$, exhibiting two twin island chains of period-$5$. 

Moreover we plot in Fig. \ref{fig1}(a) the four symmetry lines, and compute the rotation number (\ref{winding}) for $2000$ points along the symmetry line $S_1: \{ x = 0.5, -1 \le y \le 1 \}$, iterated until $n = 10^5$ [Fig. \ref{fig1}(b)]. The rotation number has a local maximum when the shear changes sign [see also the zoon in Fig. \ref{fig1}(c)], what occurs at points along the so-called shearless curve, represented as the red curve in Fig. \ref{fig1}(a), where the twist condition is also violated. 

As the system parameters change many of the invariant curves are destroyed and chaotic dynamics sets in. In nontwist systems, however, the shearless curve is remarkably resilient and survives even when neighboring curves have disappeared. Figure \ref{fig2} shows a representative example of this phenomenon, with phase portraits of the SNTM obtained for constant $b$ and varying the  parameter $a$. The twin period-$5$ islands at both sides of the shearless curve are ``dephased'', i.e. the elliptic point of one corresponds to a hyperbolic point of the other [Fig. \ref{fig2}(a)]. 

\begin{figure}
\begin{center}
\includegraphics[scale=0.1]{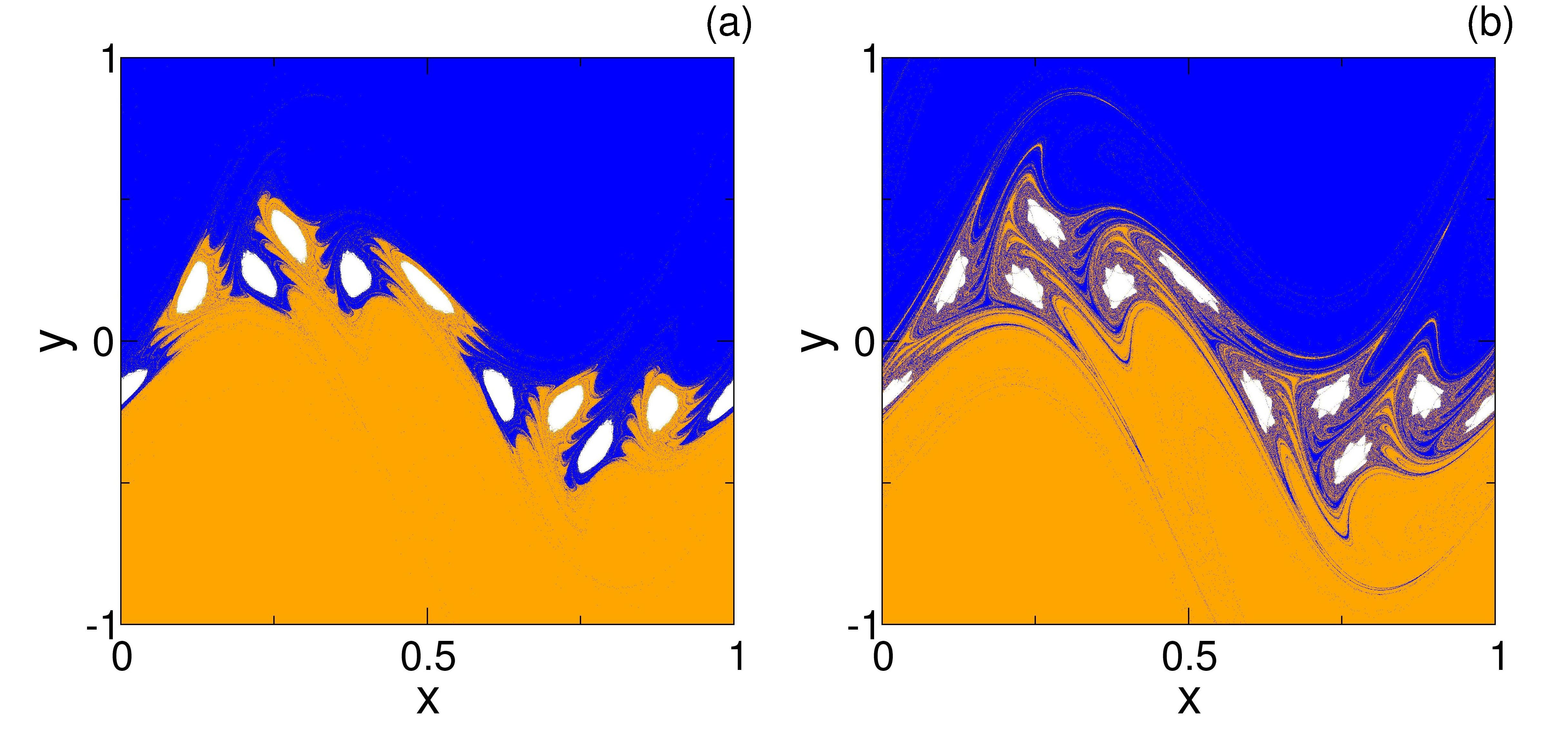}
\caption{(color online) $b = 0.619$, (a) $a = 0.642$, (b) $a = 0.645$. Blue and orange pixels represent the escape basin corresponding to the exit $y = 1$ and $-1$, respectively. Each initial condition was iterated until $n = 400$.}
\label{fig4}
\end{center}
\end{figure}

These island chains approach mutually and their separatrix reconnect, as the parameter $a$ changes (actually the reconnection affects the respective chaotic layers, since the system is no longer integrable). A further change in $a$ leaves each hyperbolic point with a homoclinic and a heteroclinic manifold and, in the region between the chains, new invariant curves appear which are not graphs over the $x$-axis and are called meanders [Fig. \ref{fig2}(b)]. Changing $a$ again makes the elliptic and hyperbolic points to collide and chaotic regions are formed, survived by the meander [Fig. \ref{fig2}(c)]. For increasing $a$ even the meander is destroyed, leaving a large chaotic region with remnants of the period-$5$ islands [Fig. \ref{fig2}(d)]. 

\begin{figure}
\begin{center}
\includegraphics[scale=0.25]{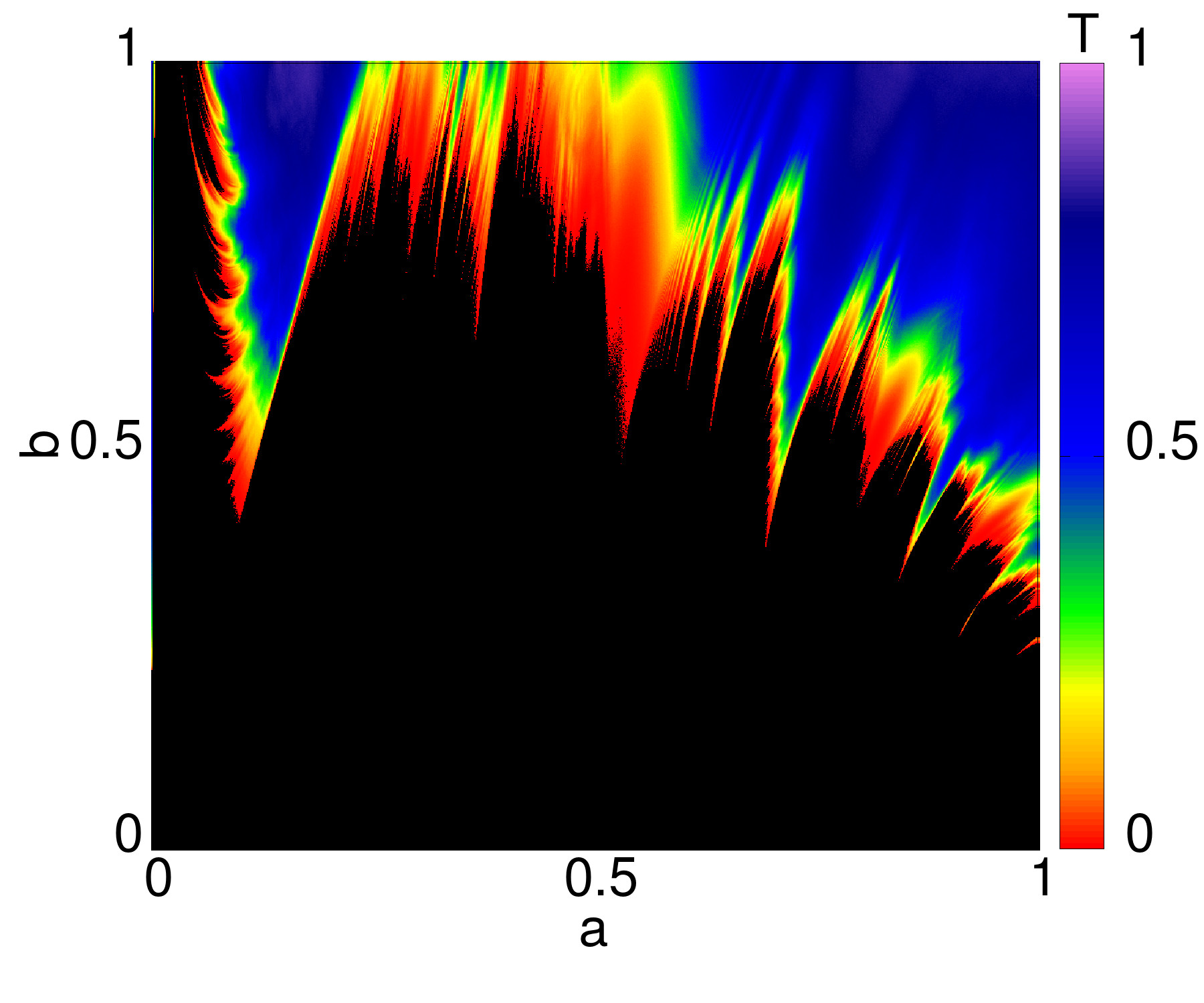}
\caption{(color online) Parameter plane ($a$ vs. $b$) obtained from the barrier transmissivity for the SNTM. We considered $10^5$ initial conditions randomly placed along the line $y = -5.0$, iterated until $n = 5000$. The colorbar indicates how many of those initial conditions reach the line $y = 5.0$.}
\label{fig5}
\end{center}
\end{figure}

While the survival of the shearless curves (or meanders) is a barrier for large scale chaotic transport, even though they disappear there is still an effective barrier, as illustrated by Fig. \ref{fig3}. In Fig. \ref{fig3}(a) and (b) we show phase portraits of the SNTM for $b = 0.619$ and $a = 0.642$ and $0.645$, respectively. In both cases after the shearless curve has disappeared and large scale transport would be possible, due to a wide chaotic region with period-$5$ island remnants. We chosen two chaotic orbits, with points painted blue and orange, with initial conditions above and below, respectively, the island remnants. For $a = 0.642$ the two chaotic orbits do not show signs of mixing, suggesting the presence of an internal transport barrier [Fig. \ref{fig3}(a)], whereas for $a = 0.645$ the two colors are mixed, signaling a higher degree of chaotic transport [Fig. \ref{fig3}(b)].

We have found that the difference between these cases is the different configuration of the unstable manifolds stemming from the Poincar\'e-Birkhoff fixed points associated with the periodic island chains in both sides of the shearless curve \cite{pre}. These unstable manifolds intercept at heteroclinic points and, if the system is nontwist, these heteroclinic points can connect the twin island chains, increasing the transport \cite{sze}. Slight variations in the system parameters, however, can alter qualitatively the geometry of the invariant manifolds and decreasing the transport. This change occurs due to the formation of structures called turnstiles \cite{mmp,rom}

Another way to regard the sudden increase of transport as the parameters are varied is to consider the SNTM as an open dynamical system, and consider that the orbits in phase space can escape to plus or minus infinity if they cross the lines $\{(x,y)|0<x<1,y=1.0 \}$ or $\{(x,y)|0<x<1, y=-1.0 \}$, respectively. We call the escape basin the set of initial conditions which produce map orbits escaping the system through a given exit. Figure \ref{fig4}, obtained for the same parameter values as Fig. \ref{fig3}, shows the escape basins of these two exits. The extent of chaotic transport is given by the degree in which these escape basins mix together. For small chaotic transport [Fig. \ref{fig3}(a)] this mixing is limited to the region neighboring the twin island chains, whereas for large chaotic transport this mixing occurs through an unbounded region of the phase space, thanks to the incursive fractal fingers [Fig. \ref{fig3}(b)]. 

The sensitive dependence of the chaotic transport on the system parameters can be quantitatively described by the transmissivity, which is the fraction of map orbits that cross the region between the twin island chain. A numerical estimate of this quantity can be obtained by placing a large number ($N = 10^5$) of initial conditions on the line $\{(x,y)|0<x<1, y=-5.0 \}$ and iterating each of them by $5 \times 10^3$ times. The transmissivity is the fraction of the orbits which reach the line  $\{(x,y)|0<x<1, y=+5.0 \}$. If this transmissivity is zero, there exists a transport barrier between the island chains, otherwise there is some degree of chaotic transport. 

In Figure \ref{fig5} we show (in a colorbar) the transmissivity of the trajectories for the SNTM as a function of its parameters $a$ and $b$. The zero transmissivity regions are painted black, indicating the existence of a robust transport barrier, which we can identify as the shearless curve (and perhaps other remaining tori in both sides of it). The boundary of the no-transmissivity region has been investigated from the point of view of a fractal curve \cite{novo}. Low transmissivity, on the other hand, can be identified with an internal transport barrier related to the presence of turnstiles, like in Figs. \ref{fig3}(a) and \ref{fig4}(a). Larger values of the transmissivity are thus characteristic of the absence of any barrier. 

\begin{figure}
\begin{center}
\includegraphics[scale=0.9]{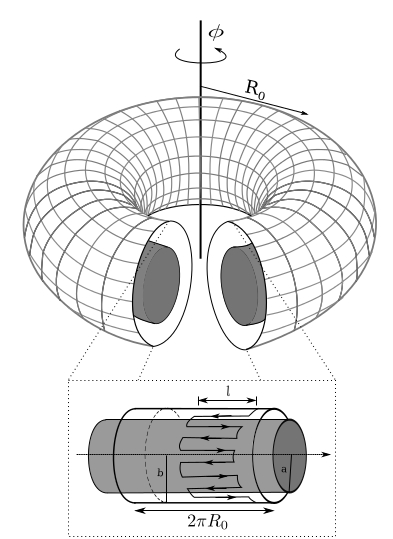}
\caption{Schematic figure of a large aspect ratio Tokamak with chaotic limiter.}
\label{fig6}
\end{center}
\end{figure}

\section{Magnetic field line maps}

Nontwist maps appear naturally in some problems of interest in Plasma Physics, like the magnetic field line structure in toroidal devices like Tokamaks and Stellarators. In this work we will focus in a Tokamak whose vessel has minor radius $b$ and major radius $R_0$ [Fig. \ref{fig6}]. The aspect ratio $R_0/b$ is supposed to be large enough that we can approximate the tokamak by a periodic cylinder of length $2\pi R_0$, in which the plasma column has a radius $a < b$. A field line point in this geometry can be identified by its cylindrical coordinates $(r,\theta,z)$, where $0 \le r < b$, $0 \le \theta < 2\pi$, and $0 \le z <2\pi R_0$. 

In the tokamak, magnetic field lines can be modeled through the following hamiltonian structure: 
 \begin{equation} 
 \frac{d\psi}{d\varphi}=-\frac{\partial H}{\partial \theta}, \qquad  
 \frac{d\theta}{d\varphi}=\frac{\partial H}{\partial \psi}.
\end{equation}
Here, $H$ is the poloidal flux, $\theta$ and $\psi$ correspond to the canonical coordinate and momentum \cite{ham1,ham2} and the toroidal angle $\varphi=z/R_0$ acts as a timelike variable. 

We can divide the Hamiltonian into two parts: $H_0$, which corresponds to the non-perturbed flux and $H_1$, the perturbed flux. Together they form  $H = H_0 + \epsilon H_1$  where:
\begin{equation}
H_0(\psi)= \int{\frac{d\psi}{q(\psi)}}, 
\end{equation}
with $q(\psi)$ as the safety profile. The conditions for MHD equilibrium imply that the magnetic field lines lie on flux surfaces $\psi = const.$, describing helical trajectories whose pitch is determined by the rotational transform $\iota(r) = 1/q(r)$.

The strength of the perturbations is given by $\epsilon$ and $H_1$ is written in terms of the following Fourier series:
\begin{equation}
H_1(\psi,\theta,\varphi)=\sum_{m,n}H_{m,n}(\psi)\cos(m\theta-n\varphi+\chi_{m,n}) ,
\end{equation}
with $m$ and $n$ as the poloidal and toroidal mode numbers and $ \chi_{m,n} $ as their phases. 

This magnetic perturbation is periodic, which allows for the creation of stroboscopic maps with sections at
 $ \varphi = \varphi_n \ = (2 \pi / s) n $,
with ($ n = 0, \pm 1, \pm 2 $) and $ s \geq 1 $. Taking ($ \psi_n , \theta_n $) as the intersection points, we can write the field line map as 
\begin{equation}
\label{tc}
(\psi_{n+1},\theta_{n+1})={\bf F}_1 (\psi_n,\theta_n). 
\end{equation}

Using $\psi=r^2/2$ as the canonical momentum \cite{ham1,ham2} and imposing that the transformation (\ref{tc}) be a canonical one, the field line map for the equilibrium part of the Hamiltonian $H_0$ is a two-dimensional symplectic map ${\bf F}_1$, derived from the magnetic field line equations, of the form
\begin{align}
 \label{rn1}
 r_{n+1} & = \frac{r_n}{1-a_1 \sin(\theta_n)}, \\
 \label{tn1}
 \theta_{n+1} & = \theta_n + \frac{2\pi}{q(r_{n+1})} + a_1 \cos(\theta_n), \qquad (\mbox{\rm mod } \, 2\pi),
\end{align}
where the parameter $a_1$ gives the toroidal correction to the cylindrical approximation. We shall use $a_1 = -0.04$. A non-monotonic safety factor profile for the equilibrium plasma is given by the expression \cite{portela}
\begin{equation}
 \label{perfilnao}
 q(r) = \frac{q_a r^2}{a^2} {\left\{
 1 - \left( 1 + \beta'\frac{r^2}{a^2} \right) 
 {\left( 1 - \frac{r^2}{a^2} \right)}^{\mu+1} H(a-r)
 \right\}}^{-1},
\end{equation}
where $\beta$, $\mu$ and $\beta'=\beta(\mu+1)/(\beta+\mu+2)$ are equilibrium parameters and $H(x)$ is the Heaviside unit-step function.

\begin{figure}
\begin{center}
\includegraphics[scale=0.08]{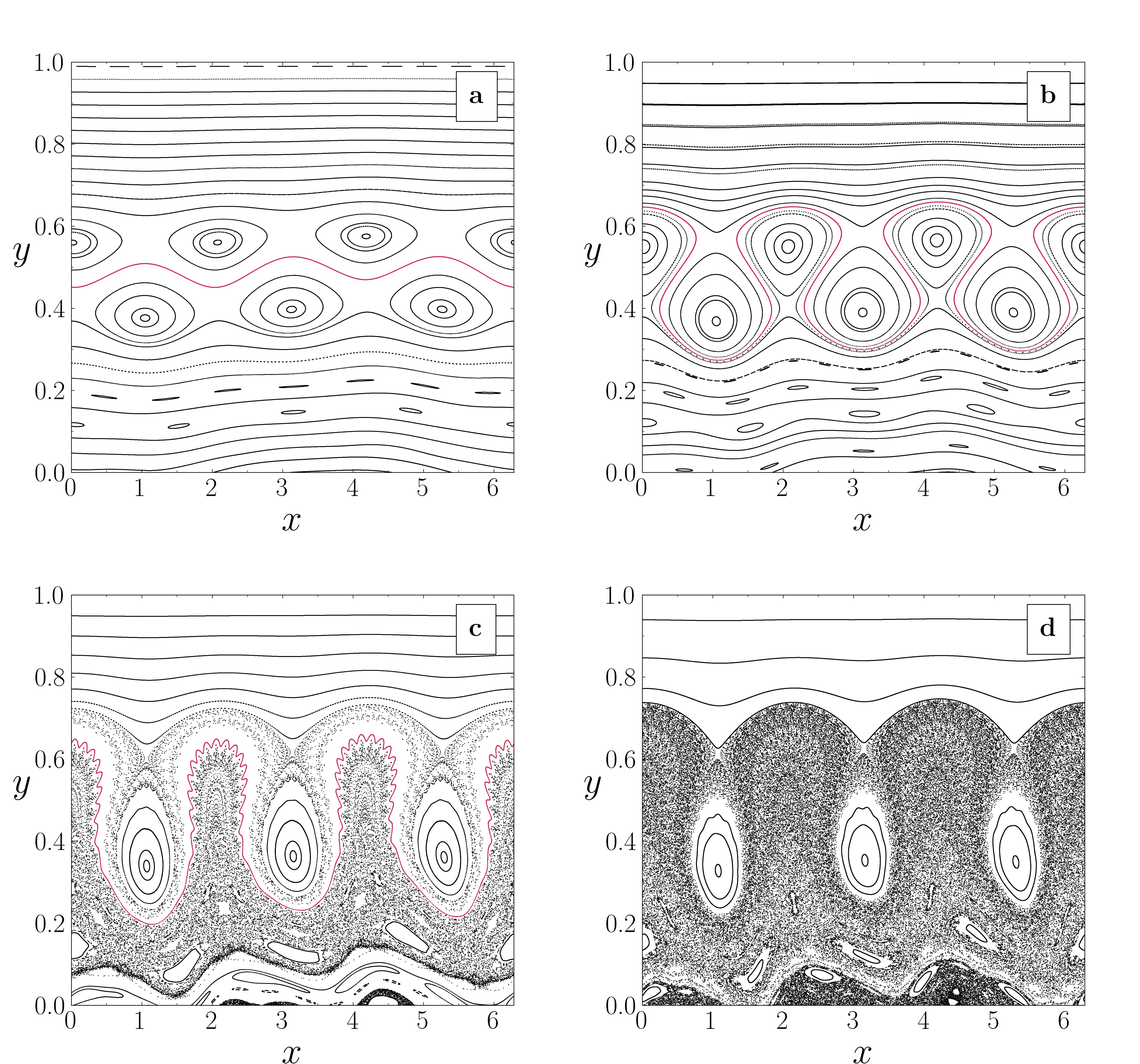}
\caption{(color online) Phase space for the Ullmann map ${\bf F}_1 \circ {\bf F}_2$ with $m=3$ and (a) $\epsilon=0.03$, (b) $0.08$, (c) $0.30$, (d) $0.40$. The remaining parameters are listed in the Appendix. The rectangular coordinates here are $x=\theta$ and $y=(b-r)/b$.}
\label{fig7}
\end{center}
\end{figure}

The conditions for the formation of a chaotic region in the Poincar\'e surface of section $(r,\theta)$ are fulfilled if a non-integrable magnetic perturbation sets in. One example is the so-called chaotic limiter, which consists in a grid of $m$ pairs of wires with length $\ell$ [Fig. \ref{fig6}], so introducing a ``time''($z$)-dependence which breaks the integrability of the equilibrium with toroidal correction given by the map ${\bf F}_1$. The magnetic field produced by a chaotic limiter with $m$ pairs of wires yields a perturbation map ${\bf F}_2$ of the form
\begin{align}
 \label{rn2}
 r_{n+1} & = r^*_{n+1} + \frac{mC\epsilon b}{m-1} {\left(\frac{r^*_{n+1}}{b}\right)}^{m-1} \sin(m\theta_{n+1}), \\
 \label{tn2}
 \theta^*_{n+1} & = \theta_{n+1} - C\epsilon {\left(\frac{r^*_{n+1}}{b}\right)}^{m-2} \cos(m\theta_{n+1}),
\end{align}
where $C = 2m\ell a^2/R_0 q_a b^2$, and $\epsilon=I_\ell/I_p$, where $I_\ell$ is the limiter current and $I_p$ is the plasma current. The composed map ${\bf F}_1 \circ {\bf F}_2$  was proposed by Ullmann and Caldas in 2000 \cite{ullmann}. The parameter values used in the numerical simulations are $b=0.21 m$ (major radius), $a = 0.18 m$ (minor radius), $\ell = 0.08 m$, $\beta = 2.0$, $\mu = 1.0$, $q_a = 3.9$ \cite{portela}. 

Phase spaces of the Ullmann map for $m=3$ are shown in Fig. \ref{fig7} for different values of the perturbation strength $\epsilon$, which is proportional to the current applied at the limiter ring. For small values of the latter we have the formation of twin dimerized islands separated by a shearless curve (in red) [Fig. \ref{fig7}(a)]. Increasing the perturbation strength these island chains approach each other and the shearless curve meanders around them [Fig. \ref{fig7}(b)]. Even when the perturbation is stronger, forming chaotic regions in both sides of the shearless curve, it continues to act as a transport barrier [Fig. \ref{fig7}(c)]. Further increase in the perturbation strength breaks down this barrier and allows a larger chaotic region with some island remnants [Fig. \ref{fig7}(d)].

\begin{figure}
\includegraphics[scale=0.5]{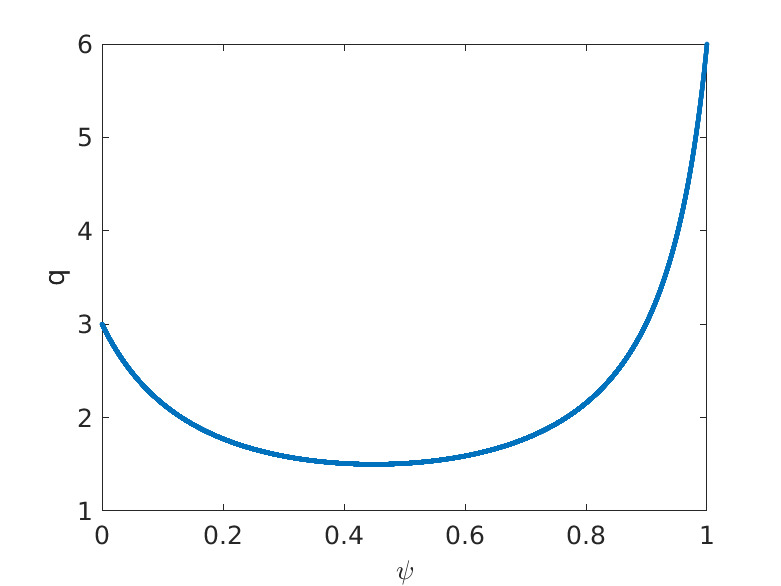} 
\caption{Non-monotonic safety factor profile of the Tokamap.}
\label{fig8}
\end{figure}

Using the original set of canonical coordinates, another symplectic field line map, the tokamap, was proposed by Balescu {\it et al.} \cite{balescu1}:
 \begin{eqnarray}
 \label{ba1}
\psi_{n+1} & = & \psi_n - \frac{\varepsilon \psi_{n+1}}{1+\psi_{n+1}} \sin(\theta_n), \\ 
 \label{ba2}
\theta_{n+1} & = & \psi_n + \frac{2\pi}{q(\psi_{n+1})} - \frac{\varepsilon \cos(\theta_k) }{(1+\psi_{k+1})^2}. 
\end{eqnarray}
The tokamap was not directly derived from the magnetic field line equations. However, it fits important characteristics for the system, namely: (i) there are no negative values of $\psi$, such that $\psi_0 = 0$ and $\psi_n \geq 0$ for all $n$; (ii) it follows a realistic safety factor profile $q(\psi)$ \cite{balescu1}. Here we use the non-monotonic profile shown in figure  \ref{fig8}, given by:
\begin{equation}
 q(\psi) = \frac{q_m}{1-\alpha(\psi-\psi_m)^2},
\end{equation}
with $\alpha=(1-q_m/q_0)\,\psi_m^{-2}$, and $\psi_m$ is the minimum of $q$ given by:
\begin{equation}
\psi_m = \left( 1 + \sqrt{\frac{1 - q_m/q_1}{1-q_m/q_0}} \right)^{-1},  
\end{equation}
where $q_0= q(0) = 3$ e $q_1=q(1) = 6$. With this profile, the mapping is also known as the \textit{revtokamap}  \cite{balescu1,wingen}. 

\begin{figure}
\includegraphics[scale=0.6]{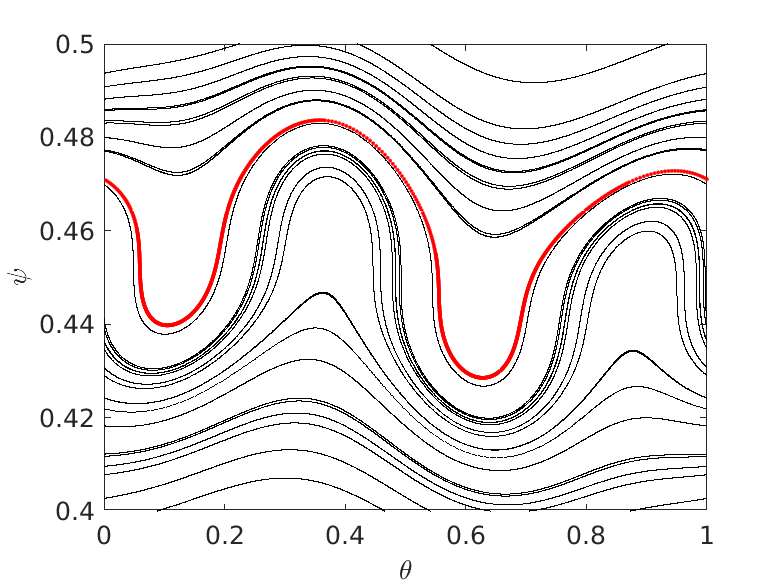} 
\caption{Poincaré section of the revtokamap (\ref{ba1})-(\ref{ba2}) for $\varepsilon=0.35$. In red we show the shearless barrier at $\psi = 0.4495$.}
\label{fig9}
\end{figure}

In Fig. \ref{fig9} we show the Poincaré section of field lines in $(\psi,\theta)$-plane for the revtokamap (\ref{ba1})-(\ref{ba2}), in which the shearless curve is drawn in red color. The qualitative evolution is similar to that exhibited by the previous maps here presented.  

\section{Shearless bifurcation and reversed current}

In the previous Section we have seen examples of symplectic field line maps, for which the existence of shearless barriers is due to the non-monotonicity of the safety factor profile of the plasma equilibrium. However, it is possible to obtain a shearless transport barrier even with monotonic safety factor profiles, provided we are close enough to some bifurcations near primary resonant islands.

Dullin, Meiss, and Sterling showed, in 2000, the existence of a shearless torus in the neighborhood of the tripling point of an elliptic fixed point of a generic Hamiltonian system \cite{dullin}. Further numerical investigations have shown the existence of shearless tori near a quadrupling bifurcation \cite{abud1}. 

\begin{figure}
\begin{center}
\includegraphics[scale=0.4]{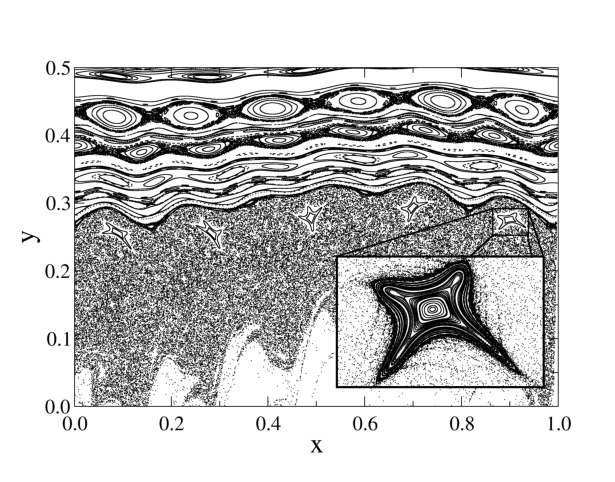}
\caption{(color online) Phase space of the Ullmann map with $\epsilon=0.1889$ and $m=6$. The inset shows a quadrupling bifurcation.}
\label{fig10}
\end{center}
\end{figure}

In the context of the Ullmann map, tripling and quadrupling bifurcations of an elliptic fixed point show up over a wide range of the perturbation parameter $\epsilon$ \cite{abud2}. In Figure \ref{fig10} we show a phase space of the Ullmann map for $\epsilon=0.189$ and $m=6$. The chaotic layer has embedded remnants of an island chain. In the inset we exhibit a period-$5$ island chain whose elliptic point bifurcated into a period-$4$ one. 

\begin{figure}
\begin{center}
\includegraphics[scale=0.6]{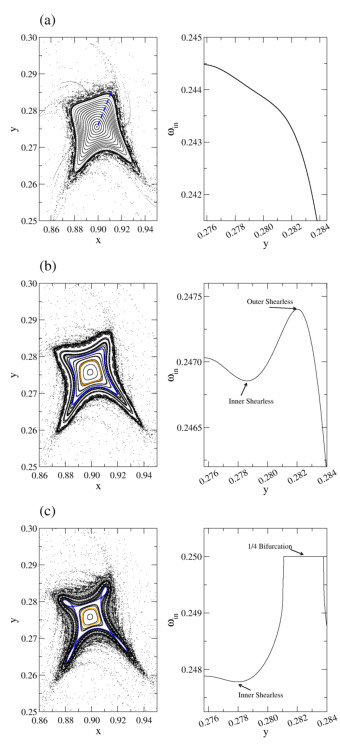}
\caption{(color online) Phase space (left) And rotation number profile (right) for the Ullmann map with $m=6$ and (a) $\epsilon = 0.185$, (b) $0.188$, (c) $0.189$.}
\label{fig11}
\end{center}
\end{figure}

In order to understand how a local shearless barrier is formed near the island chain undergoing a quadrupling bifurcation, we show in Figure \ref{fig11} the evolution of the phase spaces (left panels) and the corresponding rotation number profiles (right panels) in the neighborhood of the quadrupling bifurcation. Just before the latter [Fig. \ref{fig11}(a)] the rotation number profile is monotonic, with no local extrema. On increasing the $\epsilon$ parameter there happens a quadrupling bifurcation, through which there is a local minimum and a local maximum [Fig. \ref{fig11}(b)], and thus two shearless tori have been formed therein. As $\epsilon$ is further increased, the bump in the rotation number profile has increased its size until the local maximum achieves the value $\omega = 1/4$ yielding four stable fixed points, and the local minimum persists as well the shearless torus. 

\begin{figure}
\includegraphics[scale=0.7]{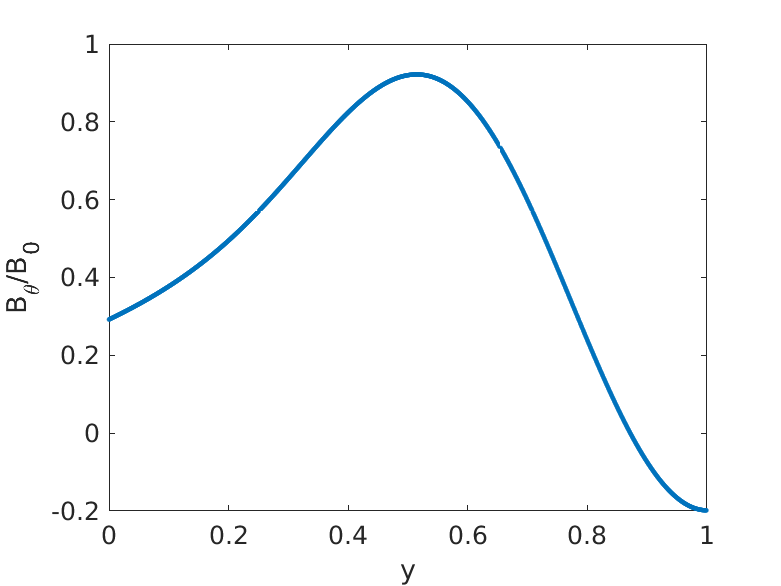} 
\caption{Poloidal magnetic field radial profile}
\label{fig12}
\end{figure}

We now discuss the dynamics of plasmas with reversed density current profile \cite{bartoloni}. Such  phenomenon has been observed in tokamak experiments \cite{fujita,stratton} and leads to magnetic fields with a non-monotonic safety factor profile. Accordingly, we consider the non-monotonic current density profile $J_z$ with a reversed current given by \cite{bartoloni}:
\begin{equation}
    J_z(r) =\frac{I_p R_0}{\pi a^2} \frac{(\delta+2)(\delta+1)}{\delta+\gamma+2}\left(1+\delta\frac{r^2}{a^2}\right)\left(1-\frac{r^2}{a^2}\right)^\gamma,
\end{equation}
where $a=0.18$ m is the plasma radius, $I_p=20$ kA is the plasma current, $\delta =-100.5$ and $\gamma=5$. Such parameters are obtained from the TCABR tokamak \cite{kroetz}. The corresponding poloidal magnetic field profile is depicted in Fig. \ref{fig12}, and the resulting safety profile is given by:
\begin{equation}
\label{qr}
q_0(r)  = q(a) \frac{r^2}{a^2}\left\{1-\left\lbrack\left(1+\beta'\frac{r^2}{a^2}\right) \left(1-\frac{r^2}{a^2}\right)^{\gamma+1}\right\rbrack \right\} \left[1-4\frac{r^2}{R_0^2}\right]^{-1/2}.    
\end{equation}
where $q(a) = 5.0$ and $\beta'=\delta(\gamma+1)/(\delta+\gamma+2)$.

\begin{figure}
\includegraphics[scale=0.7]{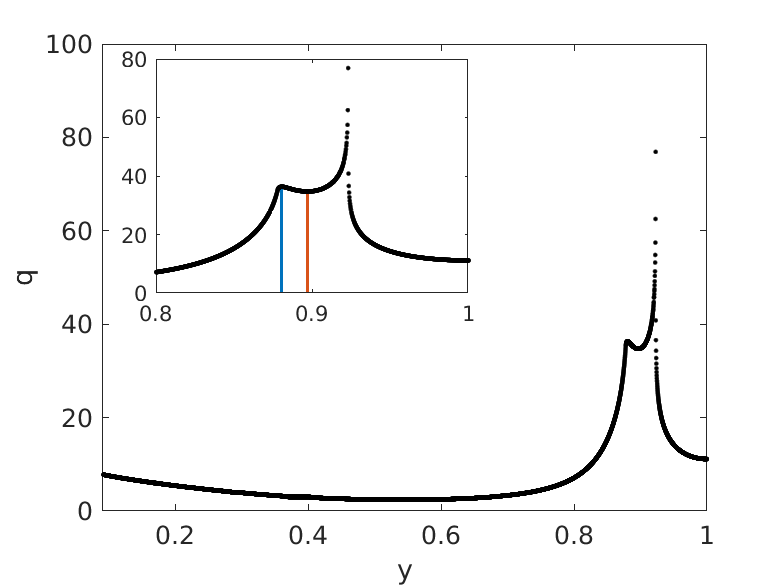} 
\caption{Numerical safety factor profile with $x=0.5$. The inset shows two shearless points: a maximum at $y_1=0.8805$ and a minimum at $y_2 = 0.8965$.}
\label{fig13}
\end{figure}

We apply this non-monotonic safety profile in the Ullmann map, with $a_1=-0.04$. On computing the rotation number we obtain Figure \ref{fig13} for a cross section at $x=0.5$. There is a divergence at $y=0.9225$ corresponding to the magnetic field reversal. 

\begin{figure}
\includegraphics[scale=0.7]{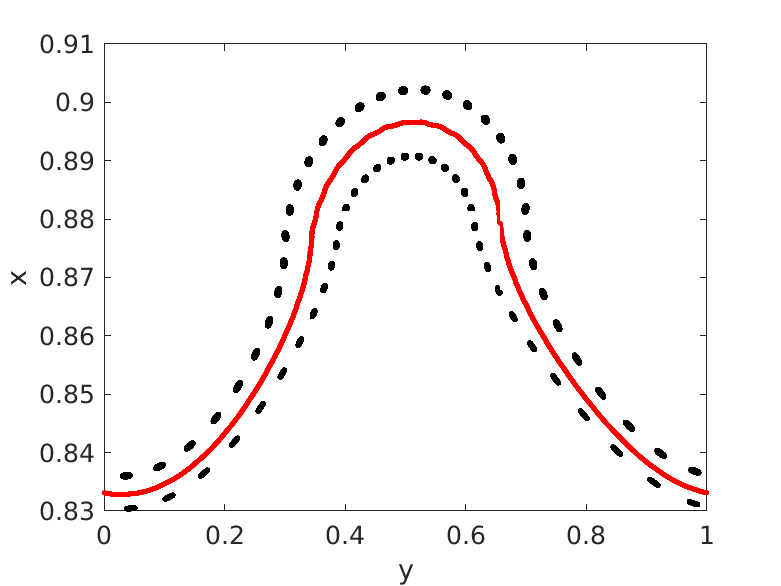} 
\caption{Shearless curve for the Ullmann map with reversed current at $y_2$ and its corresponding island chains.}
\label{fig14}
\end{figure}

The inset in Fig. \ref{fig13} shows the existence of two local extrema (one maximum at $y_1=0.8805$ and one minimum at $y_2 = 0.8965$), corresponding each to a different shearless torus. Around each of these shearless curves there are twin island chains. In figure \ref{fig14} we show the shearless curve arising at $y=y_2$ with its corresponding twin island chains. 

\section{Conclusions}

The existence of shearless barriers in tokamaks has deep consequences in terms of transport properties and the quality of plasma confinement that can be achieved. The shearless barriers are basically magnetic surfaces with some kind of robustness against symmetry-breaking perturbations. The presence of shearless barriers is usually related to non-monotonic safety factor profiles. 

A paradigm of this behavior is provided by the standard nontwist map of Morrison and del Castillo-Negrete. The shearless curve, in this case, is remarkably robust against the increase of a non-integrable perturbation strength. The location of the shearless curve is a local extremum of the rotation number profile, where the twist condition is violated for the map. 

Even after the shearless curve has been destroyed, however, transport is affected by the invariant manifold structure in the region formerly occupied by the shearless curve. The breakup of the shearless curve is extremely sensitive to the parameter values taken by the standard nontwist map. The boundary (in parameter space) between the two situation is complicated (with fractal features). 

In this paper we show two magnetic field line maps in tokamaks with non-monotonic safety factor profiles. One of them considers a tokamak with chaotic limiter, which is an external arrangement of current wires designed to create a peripheral region of chaotic field lines near the tokamak wall. For both cases the shearless tori are located at local extrema of the rotation number profile. 

We also shown that there are cases for which a field line map can exhibit shearless barriers even when the safety factor profile is monotonic. This occurs if the map parameters are close to a tripling or quadrupling bifurcation, so creating local extrema in the corresponding rotation number profiles. 

Finally, we consider explicitely a situation in which the non-monotonic safety factor profile has a well-defined physical reason, namely the existence of a current density profile with a sign reversal, a situation usually present in tokamak scenarios which partially explains why shearless barriers are so often observed. 

\begin{acknowledgments}
This work was made possible by the partial financial
support of the following Brazilian government agencies: CNPq (proc. 301019/2019-3, 428388/2018-3, 310124/2017-4), CAPES, Funda\c c\~ao Arauc\'aria, and FAPESP (grant 2018/03211-6).
\end{acknowledgments}

\end{document}